# A COMPARATIVE STUDY ON STRING MATCHING ALGORITHMS OF BIOLOGICAL SEQUENCES

**Pandiselvam.P**
Department of Computer Applications,
Ayya Nadar Janaki Ammal College,
Sivakasi, India.
*pandiselvam.pps@gmail.com*

**Marimuthu.T**
Department of Computer Applications,
Ayya Nadar Janaki Ammal College,
Sivakasi, India.
*mastersvksmca@gmail.com.*

**Lawrance. R**
Department of Computer Applications,
Ayya Nadar Janaki Ammal College,
Sivakasi, India.
*lawrancer@yahoo.com.*

*Abstract*: String matching algorithm plays the vital role in the Computational Biology. The functional and structural relationship of the biological sequence is determined by similarities on that sequence. For that, the researcher is supposed to aware of similarities on the biological sequences. Pursuing of similarity among biological sequences is an important research area of that can bring insight into the evolutionary and genetic relationships among the genes. In this paper, we have studied different kinds of string matching algorithms and observed their time and space complexities. For this study, we have assessed the performance of algorithms tested with biological sequences.

## I. INTRODUCTION

String matching is a technique to discover pattern from the specified input string. String matching algorithms are used to find the matches between the pattern and specified string. For example Let `U` is an alphabet; the basics of `U` are called symbols or characters. For example, if `U= {A, G}` then `AGAG` is a string. The pattern is denoted by `P (1…M)` the string denoted by `T (1…N)`. The pattern occurs in the string with the shifting operation.

Efficient algorithms for string matching problem can greatly aid the responsiveness of the text-editing program. String-matching algorithms are used for above problem. There are two techniques of string matching one is exact matching Needleman Wunsch (NW), Smith Waterman(SW), Knuth Morris Pratt (KMP), Dynamic Programming, Boyer Moore Horspool (BMH) and other is approximate matching (Fuzzy string searching, Rabin Karp, Brute Force).

Various string matching algorithms are used to solve the string matching problems like wide window pattern matching, approximate string matching, polymorphic string matching, string matching with minimum mismatches, prefix matching, suffix matching, similarity measure, longest common subsequence (dynamic programming algorithm), BMH, Brute Force, KMP, Quick search, Rabin Karp [12].We analyze the similarity measurements on Protein, DNA and RNA sequences by using various kinds of string matching algorithms such as Boyer Moore (BM) algorithm, NW algorithm, SW algorithm, Hamming Distance, Levenshtein Distance, Aho-Corasick (AC) algorithm, KMP algorithm, Rabin Karp algorithm, CommentZ-walter (CZW) algorithm.

In the rest of the paper, we reviewed the previous work, and then we compare the algorithms in section 3. Finally, we conclude the results in section 4.

## II. RELATED STUDY

Distance or similarity measures are essential to solve pattern recognition problems. Let us consider the following example,    String T: A C G T C G A
                              | | |
           Pattern P: _ _ _ _  C G A

Yeh.M et al [15] was uses Levenshtein distance for determining the feature vectors on the visual information such as images and videos.
Consider the example

Input sequence A:  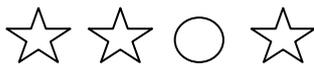

Input sequence B:  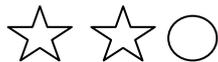

We have to find the maximum matches in A and B. For that, delete the star from the A then we get the maximum matches.

Amir.A et al [2] was proposed a new distance as the Levenshtein distance for string matching with K-Mismatches on the specified string. The proposed approach was implemented using MPI (Message Passing Interface). This algorithm is useful to establish the similarity between strings.

Knuth.D.E et al [11] was proposed a traditional pattern matching algorithm for string with running time proportional to the sum of length of strings. This algorithm named now as KMP string matching algorithm.

Hussian.I et al [9] was proposed a classical pattern matching algorithm named as Bidirectional Exact Pattern Matching algorithm (BDEPM), introduced a new idea to compare with selected text by using two pointers such as right and left simultaneously. Consider the example,
    String T: A C C G A G T
              ↑   |  |  |  ↑
    Pattern P:    C  G  A

Alsmadi.I et al [1] was evaluates two algorithms for DNA comparison those were LCS (Longest Common Substring) and LCSS (Longest Common Subsequence). They evaluate algorithms used to compute those algorithms in terms of accuracy and performance.

Input Sequence A: A ***C*** G ***T C*** G ***T***
             |     | |   |
Input sequence B: G ***C*** A ***T C*** A ***T***

The highlighted letters in the sequences are LCS of the specified two sequences, (ie) `CTCT`

Singla N et al [12] was exploiting different kinds of string matching algorithms for strings. They were decreed their preprocessing and orders that evaluate the matching. They conclude the Boyer Moore algorithm is the best for string matching.

Jain.P et al [10] was analyzed various kinds of multiple string pattern matching algorithm based on different parameters such as space, time, order to find the match and their accuracy. They were introduced a classical patter matching algorithms and the tool. They were comparing the different pattern matching algorithms with their characteristics such as total no of comparisons, shift factor, altering the order of matching. They conclude the Boyer Moore algorithm is more efficient one to apply on heterogeneous system for pattern matching.

Vidanagamachchi.S.M et al [13] was decreed two kinds of multiple pattern matching algorithms such as Aho-Corasick and CommentZ-Walter algorithms their accuracy and time taken. They were implement the code for these two algorithms and worked with peptide sequences. According to their results obtained, they conclude Aho-Corasick is performing better than CommentZ- Walter algorithm. The definition of all string matching algorithm narrated below.

*A.Hamming Distance:*

The Hamming distance is named after Richard Hamming who introduces the hamming distance for error detecting and correction codes in 1950. It measures minimum number of Substitutions required forming one DNA sequence to another. The following formula is belongs to the hamming distance.

`d`$^{HAD}$`(i,j)=∑[y`$_{i,k}$ ≠ `y`$_{j,k}$`]`    ...(1)

where,
  `d`$^{HAD}$ be the Number of dissimilarity among two sequences.
  `y`$_i$ be the first sequence, `y`$_j$ be the second sequence.
  `k` be the pairing variable.

Equation (1) used to find the minimum number of substitutions needed to form one sequence to another.

The distance itself gives the number of mismatches between the variables paired by *k*. the distance is applied to the biological sequences for finding minimum no of substitutions for changing one sequence to another.

*B. Levenshtein Distance:*

The Levenshtein distance between two strings is defined as the minimum number of edits needed to transform one string into the other, with the allowable edit operations being insertion, deletion, or substitution of a single character. It is named after Vladimir Levenshtein [2]. The following formula is belongs to the Levenshtein distance.

`d [i, j]:=minimum(d[i-1,j]+1,d[i,j-1]+1,`
                  `d[i-1,j-1]+1)`   ...(2)

where,
  `d [i,j]` be the distance matrix.
  `i, j` be the indicies.

Equation (2) is used to determine the minimum number of edits required to form one string to another.

`The Leven shtein distance between "ACGTCG" and AGGTTGA" is 3, since the following three edits change one into the other, and there is no way to do it with fewer than three edits:

1. A**G**GTTGA→A**C**GTTGA (substitution of `'C'` for `'G''`)

2. AGGT**T**GA→ACGT**C**GA  (substitution of `'T'` for `'C'`)

3. AGGTTG**A**→ACGTCG_ (deletion of `'A'` at the end).

*C. Needleman Wunsch algorithm:*

It achieves a global alignment on two sequences. It is commonly used in bioinformatics to align protein or nucleotide sequences. The NW algorithm is an example of dynamic programming, and was the first application of dynamic programming to biological sequence comparison. [12, 3].

Dynamic programming algorithm should follow the major three steps such as Initialization, Scoring matrix, Trace back. The criteria for scoring matrix is

`F`$_{ij}$`=max{F`$_{i-1,j-1}$`+S(Ai,Bj),F`$_{i-1,j}$`+d,F`$_{i,j-1}$`+d}`...(3)

where,
  `F`$_{i,j}$ be the scoring matrix.
  `i,j` be the indices of scoring matrix.
  `d` be the penalty for mismatches.
  `A`$_i$`, B`$_j$ be the elements of the sequences.
  `S(A`$_i$`,B`$_j$`)={if(Ai==Bj)` return 1;
    else return 0 }

Equation (3) is used to fill the scoring matrix in the NW algorithm.

The algorithm used to find the optimal alignment score for specified DNA sequence.

        `ATG-AG` the score: +1+1+1+0-1+1 = 3

Blocks Substitution Matrix. Scores for each position is obtained frequencies of substitutions in blocks of local alignments of protein sequences. BLOSUM-62 is appropriate for sequences of about 62% identity, while BLOSUM-80 is appropriate for **more** similar sequences.

*D. Smith Waterman Algorithm:*

It is a well-known algorithm for performing local sequence alignment; that is, for determining similar regions between two nucleotide or protein sequences [8,3]. Instead of looking at the total sequence, the Smith Waterman algorithm compares segments of all possible lengths and optimizes similarity measure. An initialization step is varying from the Needleman wunsch algorithm. The criteria for scoring matrix is

`F`$_{ij}$`=max{F`$_{i-1,j-1}$`+S(Ai,Bj),F`$_{i-1,j}$`+d,F`$_{i,j-1}$`+d}`...(4)

where,

$F_{i,j}$ be the scoring matrix.
$i,j$ be the indices of scoring matrix.
$d$ be the penalty for mismatches.
$A_i$, $B_j$ be the elements of the sequences.
$S(A_i,B_j) = \{$ if $(Ai==Bj)$ return 1;
else return -1$\}$

Equation (4) is used to fill the scoring matrix in the SW algorithm.

`ATG-AG` the score: +1+1+1+0-1+1 = 3

*E. Knuth Morris Pratt Algorithms:*

A string matching algorithm is to search the given string in the finite state machine. The KMP algorithm is the linear time string algorithm. The proper longest prefix or suffix called core.

Let `u=t [l'...r]`, `v=t [l...r]`. So that u is the longest proper prefix and suffix or core of `v` [11, 4, and 5].

To illustrate the algorithm's details, we work through a (relatively artificial) run of the algorithm, where `W = "APCDAPD"` and `S = "APC APCDAP APCDAPCDAPDE"`. At any given time, the algorithm is in a state determined by two integers:

- m which depicts the location within S which is the beginning of a perfect *match* for W.
- i the *index* in W denoting the character currently under matching.

In each step we compare S [m+i] with W[i] and advance if they are equal. This is depicted, at the start of the run.

Pattern= `ACCGTT`
string = ... `ACCGTGCGAT`

```
            | | | |
        B = "ACCG"
```

Rule fails if beginning of pattern B is in the bit we skip over.

*F. Boyer Moore algorithm:*

The BM string search algorithm is an efficient string searching algorithm that is the standard benchmark for practical string search literature. The BM algorithm is consider the most efficient string-matching algorithm in usual applications, for example, in text editors and commands substitutions.

The reason is that it woks the fastest when the alphabet is moderately sized and the pattern is relatively long.

During the testing of a possible placement of pattern P against text T, a mismatch of text character `T[i] = c` with the corresponding pattern character `P[j]` is handled as follows: If c is not contained anywhere in P, then shift the pattern P completely past `[i]`. Otherwise, shift P until an occurrence of character c in P gets aligned with `T[i]`.

As per the study the Boyer Moore algorithm is the best for the string. For example,

Let
    Input: `MNNQRKKTARPSFNMLLRAR`
    Pattern: `KKT`
After the execution of Boyer Moore algorithm
    Input: `MNNQR`***KKT***`ARPSFNMLLRAR`
                    | | |
    Pattern :       ***KKT***
    Position        ^

*G. Brute Force algorithm:*

The Brute Force algorithm compares the pattern to the text, one character at a time, until mismatching characters are found. The algorithm can be designed to stop on either the rest occurrence of the pattern, or upon reaching the end of the text [6].

Text: ABR**AKA**DABRA
Trace : AKA
    AKA
        AKA
            **AKA**

Pattern: `AKA` Input of the Brute of algorithm taken from the list of amino acids.

*H. Rabin Karp algorithms:*

The RK string searching algorithm exploits a hash function to speed up the search. The RK string searching algorithm calculates a hash value for the pattern, and for each M-character subsequence of text to be compared. If the hash values are not equal, the algorithm will estimate the hash value for next M-character sequence.

If the hash values are equal the algorithm will do the brute force comparison with the pattern and M-character sequence.
The key to RK performance is the efficient computation of hash values of the contiguous substrings of the text.

One popular and effective rolling hash function treats every substring as a number in some base, the base being usually a large prime. For example, if the substring is "AC" and the base value 1011, the hash value would be $65 \times 1011^1 + 67 \times 1011^0 = 65782$ (ASCII of 'A' is 65 and of 'C' is 67).

Rabin Karp Algorithm used in looking for similarities of two or more proteins; i.e. high sequence similarity usually implies significant structural or functional similarity.

Consider an M-character sequence as an M-digit number in base b, where b is the number of letters in the alphabet. The text subsequence `t [1...i+M-1]` is mapped into the number. The following formula is to find the subsequences of the strings.

`X(i)=t[i]*b^M-1+t[i+1]*b^M-2 +...+t[i+M-1]`
                                                    ..(5)

Equation (5) is used to determine the subsequence of the specified sequence. Furthermore, given `x(i)` we can compute `x(i+1)` for the next subsequence `t[i+1 ..i+M]` in constant time, as follows:

```
X(i+1)=t[i+1]*b^M-1+t[I+2]*b^M-2+…+t[i+M]
                                      ...(6)
```
Equation (6) depicts to find the next subsequence for the predecessor.
```
h(i)=((t[i]*b^M-1modq)+(t[i+1] *b^M-2modq)
+...+(t[i+M-1]modq))modq           ….(7)
```

where,

`x (i)` be the subsequence of the text t[i].

`h(i)` is the hash function.

`b` is the base of the string.

`q` is the prime number.

Equation (7) is used to calculate the hash value for the sub sequences. Consider the example, HbA_human:
GSAQVKGHGKKVADALTNAVAHVDDMPNALSALSDLHAHKL
G+ +VK+HGKKVA++++++AH+ D++ ++ +++LS+LH K

*I. Aho-Corasick algorithm:*

AC algorithm is a classical and suitable solution for exact string matching and it is widely used for multi pattern matching algorithm [13].

AC algorithm contains two main stages Finite state machine construction stage and matching stage. In the finite state machine construction, first construct the state machine and then regard as failure links to eliminate back tracking to root node when there is a presents of failure. In the second stage it finds out the pattern set with in the given string.

The following figure 2.1 is the result of the first stage of an AC algorithm and then we compare the string with patterns decreed in finite state machine or tire.

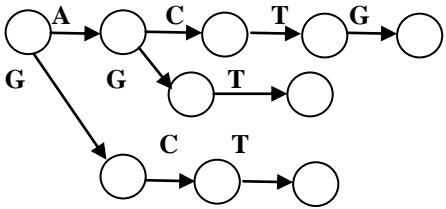

*Fig: .1 depicts the Aho-Corasick tire implementation for the patterns AC, GCT, AGT, and ACTG*

*J.CommentZ- Walter algorithm:*

CommentZ-Walter multi pattern matching algorithm combines the shifting method of BM algorithm with AC algorithm.CZW algorithm contains three stages such as finite state machine construction, shift calculation stage and matching stage. In this algorithm finite state machine or tire is constructed reverse in order to use the shifting methods of BM algorithm.

This is the result of the first stage in a CZW algorithm and then we calculate the shift calculation by using BM algorithm. Finally compare the pattern with the string or sequences.

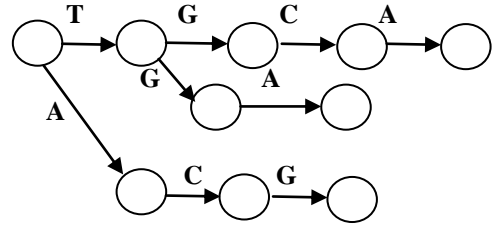

*Fig: 2. depicts the CommentZ-Walter tire implementation for patterns AC, AGT, GCA, and ACGT.*

### III. ANALYSIS OF ALGORITHMS

In this section we analyses all string matching algorithm with their accuracy and time taken to match the pattern by using online tools such as EMBOSS, GENE Wise, and manually.

| *Algorithm* | *Pre-processing* | *Execution Time* | *Accuracy* |
|---|---|---|---|
| Hamming | None | $O(N^2)$ | 81% |
| Levenshtein | None | $O(N+M)$ | 70% |
| Needleman wunsch | None | $O(MN)$ | 60% |
| Smith waterman | None | $O(MN)$ | 71.4% |
| Knuth Morris Pratt | $O(M)$ | $O(M+N)$ | 65% |
| Brute Force | None | $O(MN)$ | 66.7% |
| Boyer Moore | $O(M+N)$ | $O(MN)$ | 75% |
| Rabin Karp | $O(N)$ | $O(MN)$ | 70% |
| Aho-Corasick | None | O(N+M+Z) | 61.8% |
| CommentZ Walter | None | O(N+M+Z)+ O(MN) | 61.8% |

*Table: 1 Analyses of algorithms by using online tools and manual calculations.*

The sequence for this study taken from Genbank Accession No: JN222368 which is belongs to Marine sponge. The size of the sequence is to 1321 characters. In case of large size of sequence the process and the results won't change.

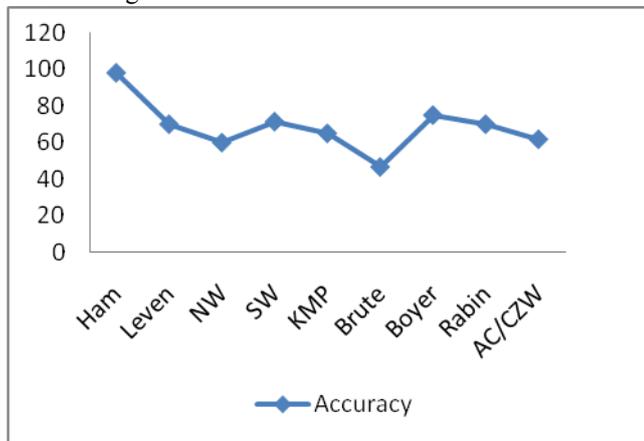

*Fig: 3 depict the accuracy analysis for various algorithms.*

## IV. CONCLUSION

In this paper, the various kinds of string matching algorithms were studied with biological sequences such as DNA and Proteins. From the studying, it is analyzed that KMP algorithm relatively easier to implement because never needs to move backwards in the input sequence, It requires extra space, Rabin Karp algorithm used to detect the plagiarism, it requires additional space for matching, Brute Force algorithm do not require preprocessing of the text or the pattern, the problem is to that its very slow, it rarely produces efficient result, Aho-Corasick algorithm is useful to multi pattern matching, CommentZ-Walter algorithm is take more time to produce the result, The Boyer Moore algorithm is extremely fast for on large sequences, it avoids lots of needless comparisons by significantly pattern relative to text, its best case running complexity is sub linear. In future we propose a speedy and efficient string matching algorithm for biological sequences.